\documentclass[amsmath,amssymb,prc,final,showpacs,twocolumn]{revtex4} 

\usepackage{bm,hyphenat,xspace} 
\usepackage{graphicx,epsfig}

\newcommand{\scl}{0.63}

\newcommand{\Ref}{Ref.}

\newcommand{\C}{{}^{12}\mathrm{C}}
\newcommand{\Cn}{{}^{13}\mathrm{C}}
\newcommand{\A}[2]{{}^{#1}\mathrm{#2}}
\newcommand{\Ox}{{}^{16}\mathrm{O}}
\newcommand{\On}{{}^{17}\mathrm{O}}

\newcommand {\mbf}[1]{{\mathbf{#1}}}
\newcommand {\vecg}[1]{\mbox{\boldmath{$#1$}} }

\begin{document}

\title {Three-body direct nuclear reactions: Nonlocal optical potential}

\author{A.~Deltuva} 
\affiliation{Centro de F\'{\i}sica Nuclear da Universidade de Lisboa, 
P-1649-003 Lisboa, Portugal }

\received{\today}


\begin{abstract}
The calculations of three-body direct nuclear reactions 
with nonlocal optical potentials are performed for the first time
using the framework of Faddeev-type scattering equations.
Important nonlocality effect is found for transfer reactions
like $d+\Ox \to p + \On$ often improving the description of the 
experimental data.
\end{abstract}

 \maketitle


Direct nuclear reactions dominated by three-body degrees of freedom
provide an important test for models of nuclear dynamics.
Extensively studied examples are deuteron $(d)$ scattering from
a stable nucleus $(A)$ and proton $(p)$ scattering from a weakly
bound nucleus $(An)$ consisting of a core $A$ and a neutron $(n)$.
The nucleon-nucleus $(NA)$ interactions employed 
in three-body calculations are usually modeled by 
optical potentials (OP) with local central and spin-orbit parts.
This local approximation yields a tremendous simplification in the
practical realization of the 
continuum discretized coupled channels (CDCC) method \cite{austern:87},
the distorted-wave Born approximation (DWBA), or various adiabatic approaches
\cite{timofeyuk:99a}
that are widely used for the description of three-body nuclear
reactions; to the best of our knowledge, the 
above mentioned methods using
nonlocal optical potentials (NLOP) have never been attempted.
However, NLOP can be included
quite easily in the framework of exact three-body Faddeev/Alt, 
Grassberger, and Sandhas (AGS) scattering theory \cite{faddeev:60a,alt:67a}
if one uses the  momentum-space representation.
Its application to three-body nuclear reactions 
\cite{deltuva:06b,deltuva:07d} has become possible recently 
due to a novel implementation of the screening and
renormalization method \cite{taylor:74a,alt:80a,deltuva:08c}
for including the long-range Coulomb force between charged particles.

 The aim of the present work is to calculate 
the observables of $d+A$ and $p+(An)$ reactions 
in a three-body model $(n,p,A)$
using NLOP and its (nearly) equivalent local optical potential (LOP)
in the framework of momentum-space AGS equations 
and thereby study the effect of the OP nonlocality.


The AGS equations \cite{alt:67a} are Faddeev-like connected-kernel 
equations that provide an exact description of the quantum three-body 
scattering problem. In contrast to the Faddeev equations \cite{faddeev:60a}
formulated for the components of the wave function, the AGS equations, 
\begin{equation}  \label{eq:Uba}
U_{\beta \alpha}(Z)  = \bar{\delta}_{\beta\alpha} \, G^{-1}_{0}(Z)  +
\sum_{\gamma}   \bar{\delta}_{\beta \gamma} \, T_{\gamma}(Z) 
\, G_{0}(Z) U_{\gamma \alpha}(Z),
\end{equation}
are a system of coupled integral equations for the transition operators 
$U_{\beta \alpha}(Z)$
whose on-shell matrix elements are scattering amplitudes and therefore
lead directly to the  observables.
In Eq.~(\ref{eq:Uba}) $ \bar{\delta}_{\beta\alpha} = 1 - \delta_{\beta\alpha}$,
$G_0(Z) = (Z-H_0)^{-1}$ is the free resolvent, and 
\begin{equation}  \label{eq:T}
T_{\gamma}(Z) = v_{\gamma} + v_{\gamma} G_0(Z) T_{\gamma}(Z)
\end{equation}
is the two-particle transition matrix with $Z=E+i0$, 
$E$ being the available three-particle energy in the 
center of mass (c.m.) system, $H_0$ the free Hamiltonian, and 
$v_{\gamma}$ the potential for the pair $\gamma$ in odd-man-out notation.
We work in momentum space where local and nonlocal potentials are treated
 in the same manner. The AGS equations are solved numerically after
partial-wave decomposition and discretization of momentum variables;
more details on employed numerical techniques can be found in 
Refs.~\cite{chmielewski:03a,deltuva:03a}.

The  AGS equations were formulated originally for a short-range potentials 
$v_{\gamma}$. Nevertheless, the method of screening and renormalization
\cite{taylor:74a,alt:80a,deltuva:08c} allows to include
the long-range Coulomb force between charged particles using the AGS framework.
The Coulomb-distorted short-range part of the
transition amplitude is obtained by solving the AGS equations with nuclear plus 
screened Coulomb potentials; the convergence of the results with the 
screening radius has to be established.
The method has been successfully applied to proton-deuteron
\cite{deltuva:05c,deltuva:05a} elastic scattering and breakup and to
three-body nuclear reactions involving deuterons or one-neutron halo nuclei
\cite{deltuva:06b,deltuva:07d}.


 For the $np$ interaction we take the realistic
CD Bonn potential \cite{machleidt:01a}, in contrast to  a simple Gaussian  
$np$ potential used in CDCC, DWBA, and adiabatic calculations.
For the hadronic part of the $NA$ interaction we take the
NLOP of Giannini and Ricco \cite{giannini} which in configuration space 
has the form
\begin{equation}  \label{eq:V}
v_{\gamma}(\mbf{r}',\mbf{r}) = H(x)[V(y) + iW(y)] + 
\vecg{\sigma}\cdot \mbf{L} H_s(x) V_s(y)
\end{equation}
with $x = |\mbf{r}'-\mbf{r}|$, $y=|\mbf{r}'+\mbf{r}|/2$, 
$H(x) = (\pi \beta^2)^{-3/2} \exp{(-x^2/\beta^2)}$,
$H_s(x) = (\pi \beta_s^2)^{-3/2} \exp{(-x^2/\beta_s^2)}$;
$V(y)$, $W(y)$, and $V_s(y)$ are the real volume, imaginary surface,
 and real spin-orbit parts, respectively, that are parametrized in the standard
way using Woods-Saxon functions. The approximately equivalent LOP is taken 
over from Ref.~\cite{giannini} as well. Both NLOP and LOP are slightly modified:
we adjust the parameter $W_N$, determining the strength of the
imaginary part, to improve the description of the experimental $NA$ 
scattering data in the considered nucleon lab energy range  from 10 to 40 MeV
 as well as the agreement 
between the $NA$ predictions of NLOP and LOP. The adjusted
values for $W_N$ are  given in Table \ref{tab:W};
the other parameters are taken from Ref.~\cite{giannini}.
In contrast to the NLOP, the LOP is energy-dependent owing to the  
equivalence transformation \cite{giannini}.
Though energy-dependent potentials were used recently \cite{deltuva:09a}
in Faddeev-type calculations, we refrain to do so in the present work.
We follow the standard procedure of fixing the  energy-dependent
parameters of the two-body OP in three-body calculations
and use two types of Hamiltonians:
(a) In $d+A$ elastic scattering both $nA$ and $pA$ OP parameters are taken
at half deuteron lab energy. 
(b) In $p+(An)$ elastic scattering  and transfer to $d+A$
 the  parameters of the $pA$ OP are
taken at the proton lab energy whereas the $nA$ potential has to be real 
in order to support an $(An)$ bound state.
Since the observables in $p+(An)$ reactions are rather insensitive to
the $nA$ potential, provided it reproduces the spectrum of bound states,
we take over the local real $nA$ potential from Ref.~\cite{deltuva:09a} 
that supports a number of bound states corresponding
to the ground and excited single-particle states of the $(An)$ nucleus,
while all Pauli-forbidden states are removed;
the potential parameters and the resulting binding energies are given
in Ref.~\cite{deltuva:09a} for  $\Cn$ and $\On$ nuclei.
The standard Hamiltonian (a) with the $nA$ potential being complex
 does not support $(An)$ bound states and therefore
does not allow for calculations of  $d+A \to p+(An)$ reactions 
while most of the available transfer data come from this type of reactions.
However, since the $d+A \to p+(An)$ and  $p+(An) \to d+A$ reactions
are related by time reversal provided the energy in the 
c.m. system is the same, we calculate the latter one using
the standard Hamiltonian (b) where the
nucleus $(An)$ can be in its ground or excited state and apply the time
reversal to obtain the observables for the former one.
This is equivalent to using the Hamiltonian (b) in the $d+A$ scattering,
a nonstandard choice.

\begin{table} [t]
\caption{\label{tab:W} 
The parameter $W_N$ of NLOP and LOP (in units of MeV) adjusted to 
the $NA$ data.} 
\begin{ruledtabular}
\begin{tabular}{l*{2}{c}}
 & $W_{N}$(NLOP)   & $W_{N}$(LOP) \\ \hline
$N$-$\C$  & 13.0 & 10.0     \\
$N$-$\Ox$ & 14.0 & 11.0     \\
$N$-$\A{40}{Ca}$ & 15.0 & 12.0    \\
\end{tabular}
\end{ruledtabular}
\end{table}

The interaction between $np$, $nA$, and $pA$ pairs is included in partial
waves with pair orbital angular momentum $L \leq 3$, 10, and 20, respectively,
and the total angular momentum is $J \leq 40$; depending on the reaction
some of these quantum numbers cutoffs can be safely chosen significantly lower,
leading, nevertheless, to well converged results.
The $pA$ channel is more demanding than the $nA$ channel due to the screened
Coulomb force, where the screening radius $R \approx 8$ to 10 fm for the
short-range part of the scattering amplitude is sufficient for convergence.

\renewcommand{\scl}{0.58}
\begin{figure}[!]
\begin{center}
\includegraphics[scale=\scl]{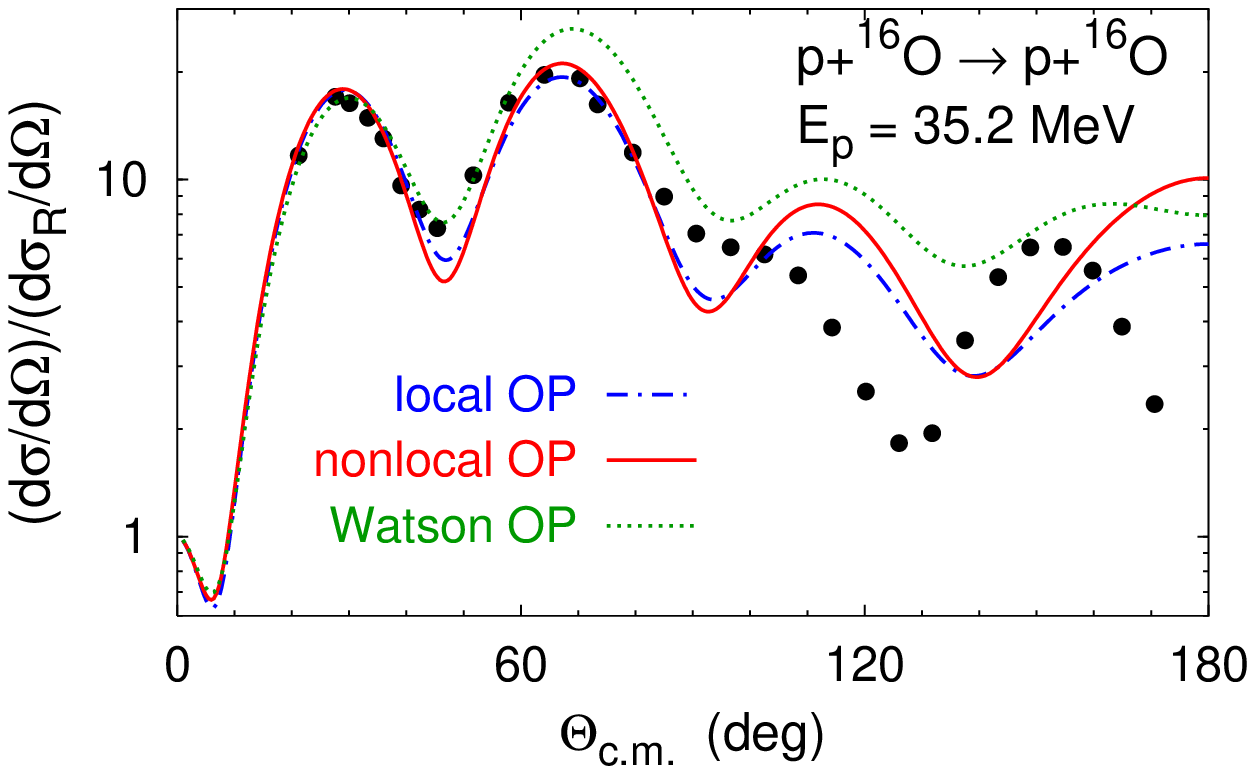}
\end{center}
\caption{\label{fig:p16O}  (Color online)
Differential cross section divided by Rutherford cross section
for  $p + \Ox$ elastic scattering at $E_p = 35.2$ MeV.
Predictions of NLOP (solid curve), LOP (dashed-dotted curve),
and Watson OP (dotted curve)  are compared with
the experimental data from Ref.~\cite{pCO35}.}
\end{figure}

In Fig.~\ref{fig:p16O} we use $p+\Ox$ elastic
scattering at proton lab energy $E_p = 35.2$ MeV as an example to illustrate
the achieved quality in fitting the 
two-body data and the approximate NLOP-LOP equivalence.
We show also the predictions of the OP by Watson {\it et al.} \cite{watson}
 to demonstrate that the adjusted NLOP and LOP describe the
$NA$ data at least as good as the traditionally used potentials.
An agreement of similar quality is found in all considered cases.

\begin{figure}[!]
\begin{center}
\includegraphics[scale=\scl]{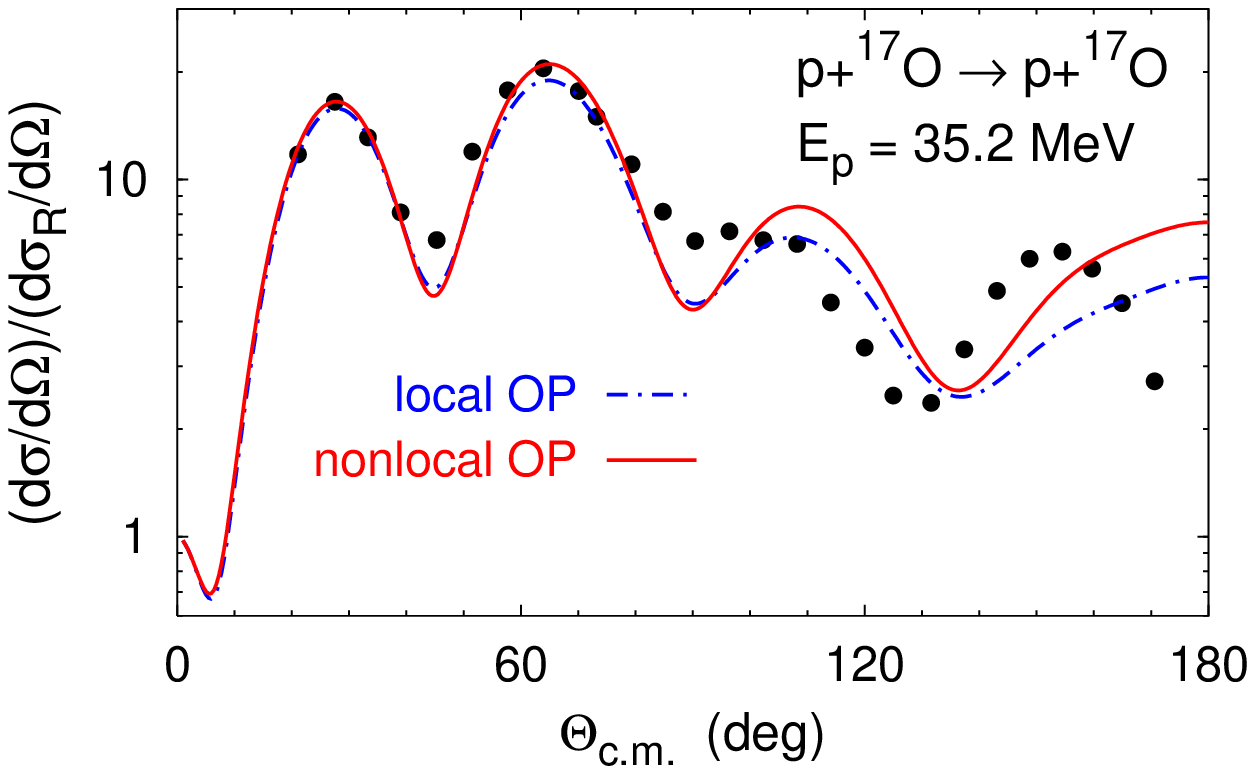}
\end{center}
\caption{\label{fig:p17O}  (Color online)
Differential cross section divided by Rutherford cross section
for  $p + \On$ elastic scattering at $E_p = 35.2$ MeV.
Predictions of NLOP (solid curve) and LOP (dashed-dotted curve)
are compared with the experimental data from Ref.~\cite{pCO35}.}
\end{figure}

\begin{figure}[!]
\begin{center}
\includegraphics[scale=\scl]{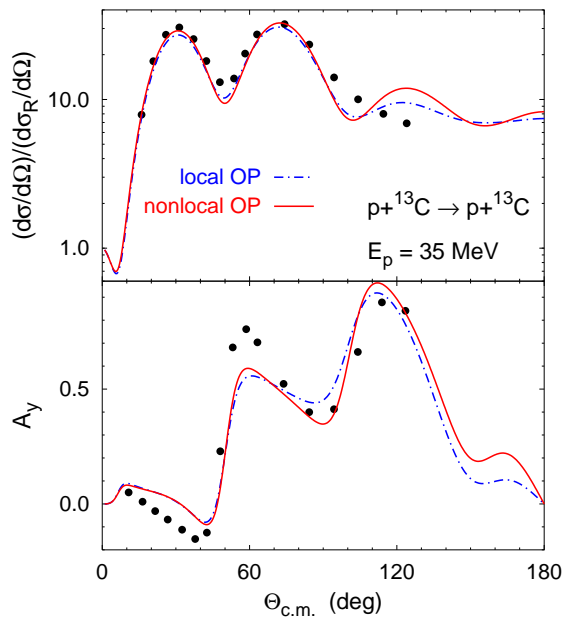}
\end{center}
\caption{\label{fig:pC35}  (Color online)
Differential cross section divided by Rutherford cross section
and proton analyzing power for 
for  $p + \Cn$ elastic scattering at $E_p = 35$ MeV.
Curves as in Fig.~\ref{fig:p17O} and the experimental data are from 
Ref.~\cite{pC35}.}
\end{figure}

In Figs.~\ref{fig:p17O} and \ref{fig:pC35}
we show the results of the three-body calculation
for elastic proton scattering from $\On$ and $\Cn$ nuclei
around $E_p = 35$ MeV.
 The mutual agreement between  NLOP and LOP predictions and 
with the data is as good as in the two-body case shown in Fig.~\ref{fig:p16O}.
This is not surprising
since the differential cross section and the proton analyzing power 
in $p+(An)$ elastic scattering are known to
correlate strongly  with the corresponding observables
in $p+A$ elastic scattering.
 Therefore, the effect on the OP nonlocality  is tiny;
 even the small differences between NLOP and LOP predictions seen in
$p+\Ox$ cross section at large angles are reproduced well
in  $p+\On$ elastic scattering.

\begin{figure}[!]
\begin{center}
\includegraphics[scale=\scl]{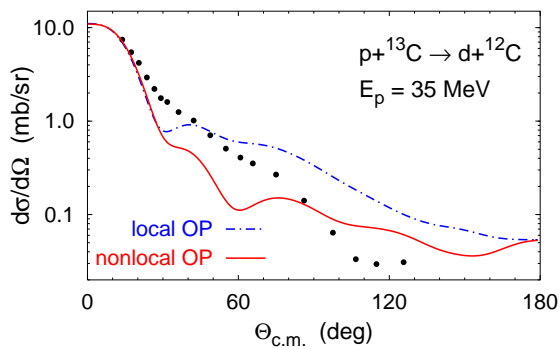}
\end{center}
\caption{\label{fig:pCd}  (Color online)
Differential cross section for  $p + \Cn \to d+\C$ transfer 
at $E_p = 35$ MeV. Curves as in Fig.~\ref{fig:p17O} 
and the experimental data are from Ref.~\cite{pC35d}.}
\end{figure}

\begin{figure}[!]
\begin{center}
\includegraphics[scale=\scl]{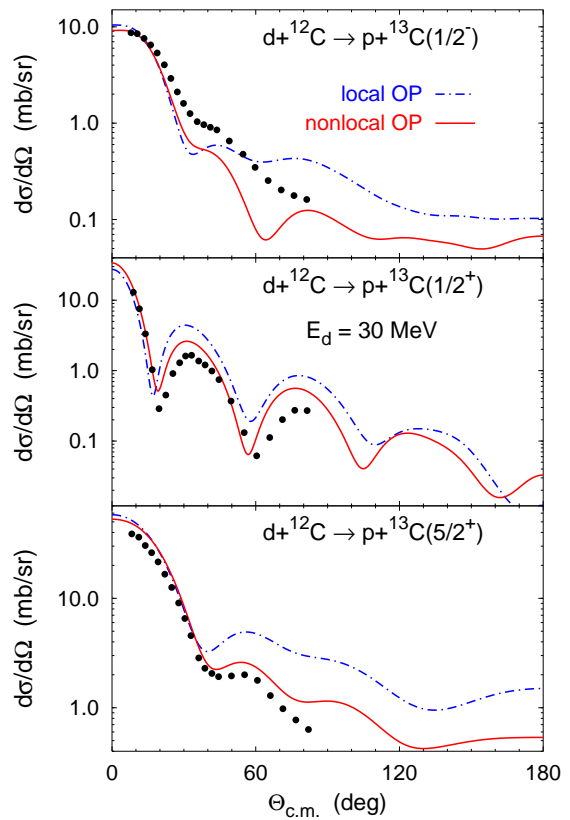}
\end{center}
\caption{\label{fig:dC30p}  (Color online)
Differential cross section for  $d + \C \to p+\Cn$ transfer to 
ground and excited $\Cn$ states at $E_d = 30$ MeV.
Curves as in Fig.~\ref{fig:p17O} and the data are from \Ref~\cite{dC30p}.}
\end{figure}

The Hamiltonian (b) used to calculate  $p + \Cn$ elastic scattering
in Fig.~\ref{fig:pC35} allows for the transfer to $d+\C$ as well;
the AGS equations for both reactions are solved simultaneously. 
In contrast to elastic scattering,
the effect of the OP nonlocality becomes significant in the 
$p+\Cn \to d+\C$ transfer cross section at angles
$\Theta_{c.m.} > 30$ deg as demonstrated in Fig.~\ref{fig:pCd},
although it does not improve the description of the experimental data.
We therefore expect the OP nonlocality to be important also 
in the inverse reactions, i.e., $d+A \to p+(An)$, where more data exist
with the final nucleus
$(An)$ being in its ground or excited state; the observables 
are calculated using the Hamiltonian (b) as described above.
We start in Fig.~\ref{fig:dC30p}
with the $d+\C \to p+\Cn$ reaction at deuteron lab energy $E_d=30$ MeV
which in the  $p+\Cn \to d+\C$ case corresponds to $E_p = 30.6$,  27.3, 
and 26.5 MeV for $\Cn$ states $1/2^-$, $1/2^+$, and $5/2^+$, respectively.
For the transfer to the $\Cn$ ground state $1/2^-$
the shape of the experimental data and theoretical predictions are
similar to the ones of $p+\Cn \to d+\C$ reaction at $E_p = 35$ MeV 
in Fig.~\ref{fig:pCd} as expected from the detailed balance,
given the small difference in energy.
Thus, except for forward angles,
the transfer reactions involving  the $\Cn$ ground state $1/2^-$
are described rather unsuccessfully much like it was found in 
Ref.~\cite{deltuva:09a}.
In the case of the transfer to $\Cn$ excited states  $1/2^+$ and $5/2^+$ 
the  OP nonlocality is again important, and the predictions of the NLOP
account for the data quite successfully, in contrast to the ones of the LOP 
and Ref.~\cite{deltuva:09a}.  The differential cross section for the
transfer to $\Cn$ $1/2^+$ state is increased by the NLOP 
at forward angles and decreased at  $\Theta_{c.m.} > 20$ deg 
while for the $5/2^+$ state it is significantly decreased at  
$\Theta_{c.m.} > 35$ deg such that the data is slightly overestimated by almost
 a constant factor that may be associated with the spectroscopic factor.

\begin{figure}[!]
\begin{center}
\includegraphics[scale=\scl]{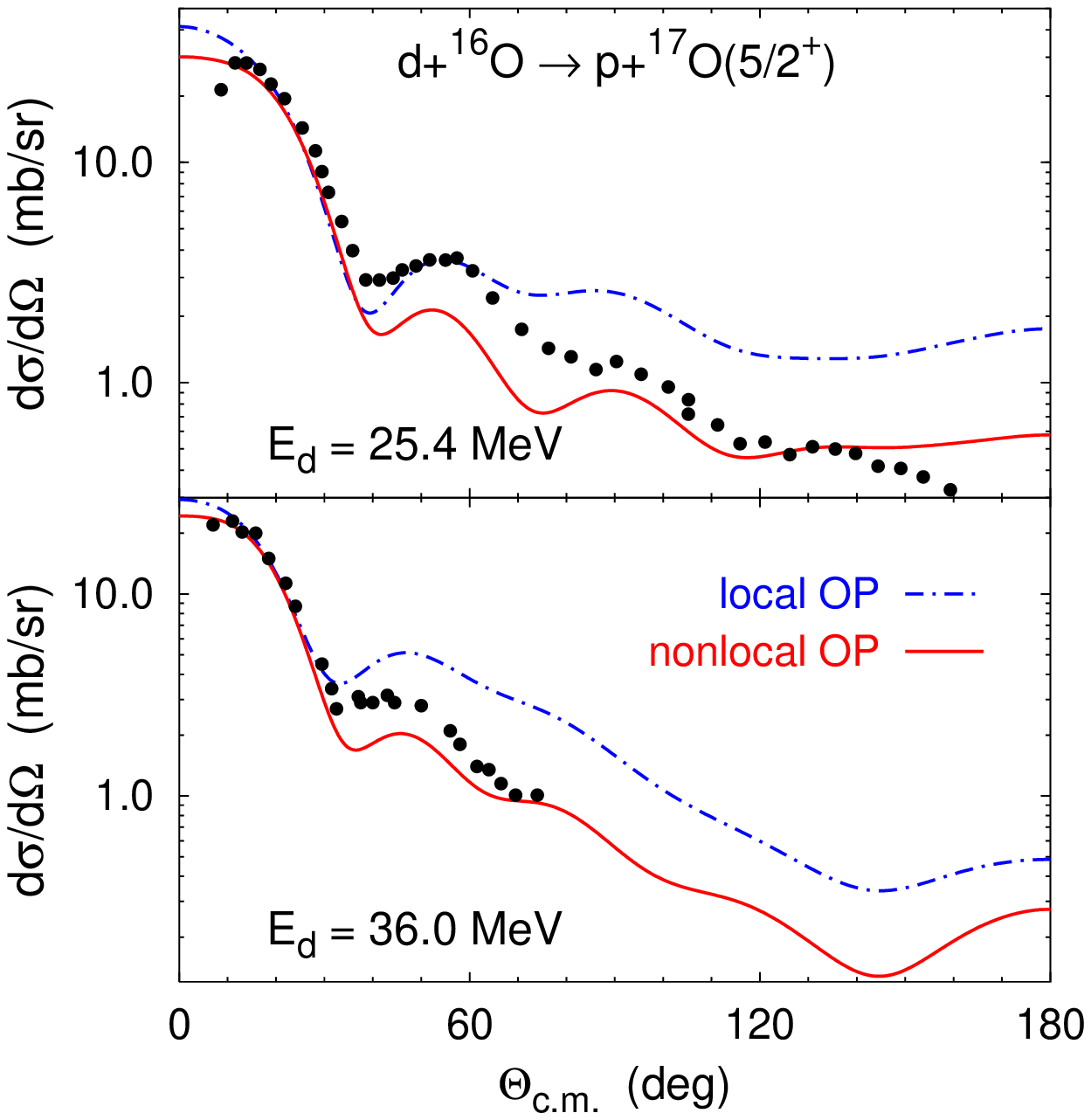}
\end{center}
\caption{\label{fig:dOpd}  (Color online)
Differential cross section for  $d + \Ox \to p+\On$ transfer to 
the ground state $5/2^+$ of $\On$ at $E_d = 25.4$ and 36.0 MeV.
Curves as in Fig.~\ref{fig:p17O}.
The experimental data are from \Ref~\cite{dO25-63}.}
\end{figure}

\begin{figure}[!]
\begin{center}
\includegraphics[scale=\scl]{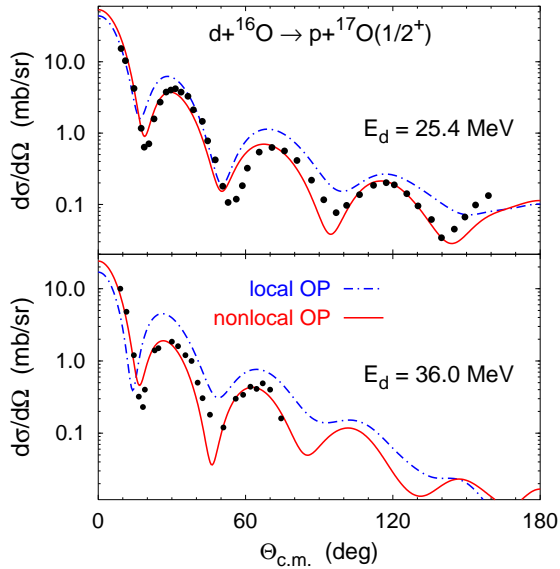}
\end{center}
\caption{\label{fig:dOps}  (Color online)
Differential cross section for  $d + \Ox \to p+\On$ transfer to 
the excited state $1/2^+$ of $\On$ at $E_d = 25.4$ and 36.0 MeV.
Curves as in Fig.~\ref{fig:p17O}.
The experimental data are from Ref.~\cite{dO25-63}.}
\end{figure}

The results for the $d+\Ox \to p + \On$ transfer cross sections
at $E_d = 25.4$ and 36.0 MeV are presented in Figs.~\ref{fig:dOpd}
and \ref{fig:dOps}. The corresponding proton lab energy in the
inverse reactions $p+\On \to d + \Ox$ is
$E_p = 25.9$ and 35.9 MeV for $\On$ ground state $5/2^+$
and $E_p = 25.0$ and 35.0 MeV for $\On$ excited  state $1/2^+$, respectively.
The effect of the OP nonlocality is again sizable and qualitatively
very similar to the one observed in $d+\C \to p + \Cn$ reactions 
involving $\Cn$ excited states $5/2^+$ and $1/2^+$.
The differential cross section for the
transfer to the $\On$ ground state $5/2^+$ is decreased at very forward angles
and at $\Theta_{c.m.} > 30$ deg. Although in particular narrow angular 
regimes the predictions of the LOP are closer to the data, the NLOP provides
a better overall description of the data, especially of its shape.
In the case of the transfer to the $\On$ excited state $1/2^+$
the NLOP reproduces the data almost perfectly.

\renewcommand{\scl}{0.54}
\begin{figure}[t]
\begin{center}
\includegraphics[scale=\scl]{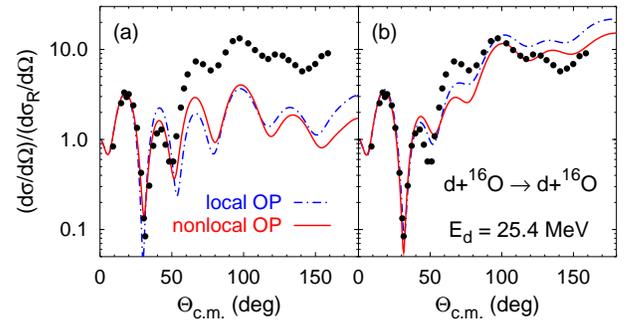} 
\end{center}
\caption{\label{fig:dO25}
Differential cross section divided by Rutherford cross section for
$d+\Ox$ elastic scattering at $E_d = 25.4$ MeV.  Results obtained with
the Hamiltonian of type (a) and (b) are shown on the left and right side,
respectively. Curves as in Fig.~\ref{fig:p17O}.
The experimental data are from Ref.~\cite{dO25-63}.}
\end{figure}

\renewcommand{\scl}{0.5}
\begin{figure}[!]
\begin{center}
\includegraphics[scale=\scl]{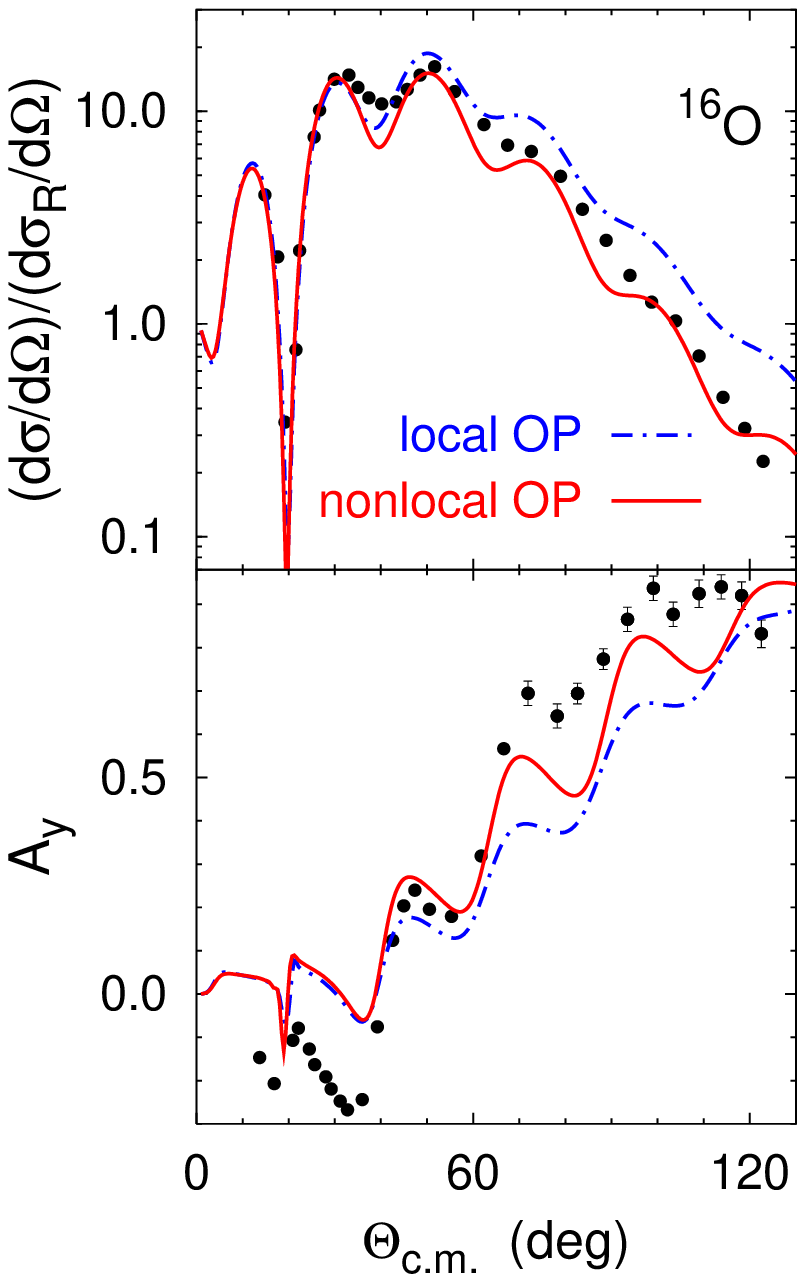} 
\includegraphics[scale=\scl]{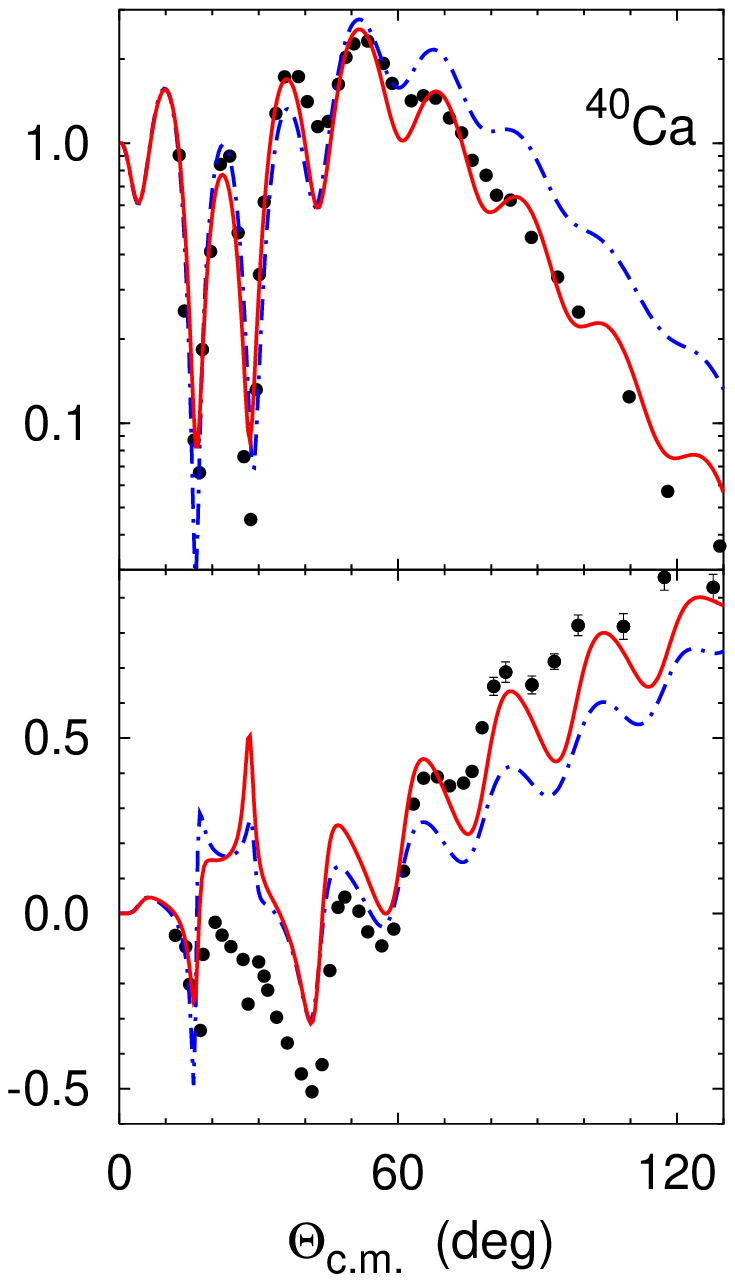} 
\end{center}
\caption{\label{fig:dOCa}
Differential cross section divided by Rutherford cross section
and deuteron vector analyzing power for  deuteron elastic scattering
from $\A{16}{O}$ and $\A{40}{Ca}$ nuclei at $E_d = 56$ MeV.
Curves as in Fig.~\ref{fig:p17O}.
The experimental data are from Ref.~\cite{matsuoka:86}.}
\end{figure}

Next we consider deuteron-nucleus elastic scattering 
that is usually described using the Hamiltonian of type (a). 
We show in Fig.~\ref{fig:dO25} the results for $d+\Ox$
elastic cross section at $E_d = 25.4$ MeV.
The effect of the OP nonlocality is visible at larger angles
 $\Theta_{c.m.} > 30$ deg and moves the predictions towards the data
up to $\Theta_{c.m.} = 60$ deg; beyond 60 deg the data is strongly
underpredicted by both NLOP and LOP results. This indicates that
the Hamiltonian (a) is too absorptive in the considered case
where negative energies in the two-body $NA$ subsystem have a large
weight in the three-body scattering equations as discussed in 
Ref.~\cite{deltuva:09a}; it was found there that using energy-dependent
potential that becomes real and therefore less absorptive at negative 
$NA$ energies significantly increases the cross section at 
$\Theta_{c.m.} > 60$ deg.
For curiosity in Fig.~\ref{fig:dO25} we present also results of
the Hamiltonian (b) which is less absorptive since the $nA$ potential
is real. Surprisingly, with this choice both the NLOP and the LOP 
roughly account for the data also at large angles.
Thus, NLOP and LOP with the  Hamiltonian of type (b) describe elastic
$d+\Ox$ and $p+\On$ data with similar quality while for the transfer
reactions the NLOP is clearly superior.

Finally in Fig.~\ref{fig:dOCa} we present observables for
$d+\Ox$ and $d+\A{40}{Ca}$ elastic scattering at $E_d = 56$ MeV
calculated with the  Hamiltonian of type (a).
Visible differences between NLOP and LOP show up at large angles
$\Theta_{c.m.} > 60$ deg. The predictions of NLOP are closer to the data
but show too sharp oscillations; a similar feature is present
already in the $p+A$ and $p+(An)$ observables in 
Figs.~\ref{fig:p16O} --- \ref{fig:pC35} but is less pronounced there.
Both NLOP and LOP fail in reproducing small-angle deuteron vector analyzing
power $A_y$ much like other potentials as found in Ref.~\cite{deltuva:09b}.
The effect of the OP nonlocality is of comparable size also for deuteron
 tensor analyzing powers, but it is unable to resolve the discrepancy 
\cite{deltuva:09b} in the large-angle $A_{xz}$.
The results for deuteron-$\C$ elastic scattering exhibit similar features
as those for deuteron-$\Ox$ and are not shown separately.

To be sure that the observed OP nonlocality effect is not a consequence
of only approximate  NLOP-LOP equivalence,
we performed the following test calculations.
We varied the parameter $W_N$ by $\pm 1$ MeV thereby inducing changes in the
$NA$ observables of a size comparable to the NLOP-LOP difference.
As a consequence, similar changes occur in $p+(An)$ elastic observables,
but for $d+A$ elastic scattering and especially for all transfer reactions
the induced changes are considerably smaller than the observed 
NLOP-LOP difference. Thus, the imperfection in the  NLOP-LOP equivalence 
does not affect our conclusions on the importance of the OP nonlocality.
A possible reason for an overall better description of the 
three-body observables by NLOP may be that NLOP fits the
$NA$ data over a broader energy range compared to LOP which, although 
being energy-dependent, has to be chosen at a fixed energy.

In summary, we performed the calculations of three-body direct nuclear 
reactions with the NLOP for the first time. Exact scattering equations in
the AGS form were solved and the Coulomb interaction was included using
the method of screening and renormalization.
The OP nonlocality effect is found to be very small for $p+(An)$ elastic 
scattering, moderate for $d+A$ elastic scattering at larger angles,
and especially important for transfer reactions $p+(An) \to d+A$ 
and $d+A \to p+(An)$. In the latter case the NLOP is clearly more successful 
in accounting for the data in transfer reactions involving
$\Cn$ and $\On$ nuclei in the states $1/2^+$ and $5/2^+$.
We hope that the present work, demonstrating the feasibility of the 
calculations with NLOP and the importance of the nonlocality, 
 will stimulate the development of new and
more precise nonlocal nuclear interaction models.

\vspace{1mm}
The author thanks A.~C.~Fonseca for comments on the manuscript.
The work is supported by the Funda\c{c}\~{a}o para a Ci\^{e}ncia
e a Tecnologia (FCT) grant SFRH/BPD/34628/2007.



\end{document}